\documentclass[aps, prb, reprint,amsmath,amssymbols,superscriptaddress]{revtex4-1}

\usepackage{graphicx}
\usepackage{bm}
\usepackage{color}
\usepackage{soul}
\usepackage{multirow,hhline}
\usepackage{lipsum}

\begin{document}


\title{Ginzburg-Landau theory of the superheating field anisotropy of layered superconductors}

\author{Danilo B. Liarte}
\email[]{dl778@cornell.edu}
\affiliation{Laboratory of Atomic and Solid State Physics, Cornell University, Ithaca, NY, USA}
\author{Mark K. Transtrum}
\affiliation{Department of Physics and Astronomy, Brigham Young University, Provo, Utah 84602, USA}
\author{James P. Sethna}
\email[]{sethna@lassp.cornell.edu}
\affiliation{Laboratory of Atomic and Solid State Physics, Cornell University, Ithaca, NY, USA}




\date{\today}

\begin{abstract}
We investigate the effects of material anisotropy on the superheating field of layered superconductors. We provide an intuitive argument both for the existence of a superheating field, and its dependence on anisotropy, for $\kappa = \lambda / \xi$ (the ratio of magnetic to superconducting healing lengths) both large and small. On the one hand, the combination of our estimates with published results using a two-gap model for MgB${}_2$ suggests high anisotropy of the superheating field near zero temperature. On the other hand, within Ginzburg-Landau theory for a single gap, we see that the superheating field shows significant anisotropy only when the crystal anisotropy is large and the Ginzburg-Landau parameter $\kappa$ is small. We then conclude that only small anisotropies in the superheating field are expected for typical unconventional superconductors near the critical temperature. Using a generalized form of Ginzburg Landau theory, we do a quantitative calculation for the anisotropic superheating field by mapping the problem to the isotropic case, and present a phase diagram in terms of anisotropy and $\kappa$, showing type I, type II, or mixed behavior (within Ginzburg-Landau theory), and regions where each asymptotic solution is expected. We estimate anisotropies for a number of different materials, and discuss the importance of these results for radio-frequency cavities for particle accelerators.

\end{abstract}


\maketitle


\section{Introduction}
\label{sec:introduction}

A superconductor in a magnetic field parallel to its surface can be metastable to flux penetration up to a (mis-named) \emph{superheating field} $H_{\text{sh}}$, which is above the field at which magnetism would penetrate in equilibrium ($H_{\text{sh}}>H_c$ and $H_{\text{sh}}>H_{c1}$ for type-I and type-II superconductors, respectively). Radio-frequency cavities used in current particle accelerators routinely operate in this metastable regime, which has prompted recent attention on theoretical calculations of this superheating field~\cite{catelani08, transtrum11}. The first experimental observation of the superheating field dates back to 1952~\cite{garfunkel52}, and a quantitative description has been given early by Ginzburg in the context of Ginzburg-Landau (GL) theory~\cite{ginzburg58}. Since then, there have been many calculations of the superheating field within the realm of GL~\cite{kramer68, gennes65, galaiko66, kramer73, fink69, christiansen69, chapman95, dolgert96}. In particular, Transtrum et al.~\cite{transtrum11} studied the dependence of the superheating field on the GL parameter $\kappa$. Here we use their results and simple re-scaling arguments to study the effects of material anisotropy in the superheating field of layered superconductors.

The layered structure of many unconventional superconductors is not only linked with the usual high critical temperatures of these materials; it also turned small corrections from anisotropy effects into dominant properties~\cite{tinkham96}. For instance, the critical current of polycrystalline magnesium diboride is known to vanish far below the upper critical field, presumably due to anisotropy of the grains (the boron layers inside each grain start superconducting at different temperatures, depending on the angle between the grain layers and the external field)~\cite{patnaik01, eisterer03}. Cuprates, such as BSCCO, exhibit even more striking anisotropy, with the upper critical field varying by two orders of magnitude depending on the orientation of the crystal with respect to the direction of the applied magnetic field~\cite{tinkham96}.

One would expect that such anisotropic crystals also display strong anisotropy on the superheating field. Here we show that this is typically not true near the critical temperature. Type II superconductors, which often display strong anisotropic properties, also have a large ratio between penetration depth and coherence length (the GL parameter $\kappa$), which, as we shall see, considerably limits the effects of the Fermi surface anisotropy on the superheating field. At low temperatures, heuristic arguments suggest that crystal anisotropy might be important for the superheating field of multi-band superconductors, such as MgB${}_2$ (section~\ref{sec:mgb2}).

It is usually convenient to characterize crystalline anisotropy by the ratio of the important length scales of superconductors, within Ginzburg-Landau theory,
\begin{eqnarray}
\gamma = \frac{\lambda_c }{ \lambda_a } = \frac{ \xi_a }{ \xi_c } = \sqrt{\frac{m_c}{m_a}},
\label{eq:anisotropy}
\end{eqnarray}
where $\lambda$ is the penetration depth, $\xi$ is the coherence length, $m$ is the effective mass, and the indices $c$ and $a$ are associated with the layer-normal axis $\bm{c}$, and an in-plane axis, respectively. Note that $\lambda_i$ is associated with the screening by supercurrents flowing along the $i$-th axis~\cite{tinkham96}. Hence for a magnetic field parallel to a flat surface of superconductor, $\lambda=\lambda_c$ only when $\bm{c}$ is perpendicular to both the magnetic field and the surface normal; counterintuitively, $\lambda=\lambda_a$ for $\bm{c}$ parallel to the magnetic field or the surface normal. In this paper, we show that the anisotropy of $H_{\text{sh}}$ is larger for larger $\gamma$ and smaller $\kappa_\parallel$, and behaves asymptotically as: $H_{\text{sh}}^\parallel / H_{\text{sh}}^\perp \approx 1$ for $\kappa_\parallel \gg 1 / \gamma$, and $H_{\text{sh}}^\parallel / H_{\text{sh}}^\perp \approx \gamma^{1/2}$, for $\kappa_\parallel \ll 1$. We begin with two simple qualitative calculations that motivate the two limiting regimes intuitively. We shall then turn to the full GL calculation, which we map, using a suitable change of variables and rescaling of the vector potential, onto an isotropic free energy, and discuss the implications of these results for several materials. We then discuss a generalization of our simple estimates for MgB${}_2$ at lower temperatures, using results from a two-gap model, and make some concluding remarks.

\section{Simple estimations of $H_{\text{sh}}$ in the large and small-$\kappa$ regimes}
\label{sec:estimations}

In this section, we discuss two simple arguments to motivate and estimate
the superheating field for both isotropic and anisotropic superconductors.
These complement the systematic calculation within Ginzburg-Landau theory 
presented in section~\ref{sec:gl}. Our first estimate applies both to small
and large $\kappa$ superconductors; for large $\kappa$, it discusses the initial
entry of the core of a vortex into the superconductor. The second estimate 
(for large $\kappa$, generalizing Bean and Livingston~\cite{bean64}), discusses
the field needed to push the core from near the surface into the bulk of the 
superconductor, fighting the attraction of the vortex to the surface.
Both methods yield estimates for the superheating field that
are compatible, up to an overall factor, with the estimates of anisotropic
Ginzburg-Landau theory of section~\ref{sec:gl}. However, we shall discuss
qualitative differences between the sinusoidal modulations at $H_{\text{sh}}$
predicted by linear stability theory and the unsmeared vortices used in
these two simple pictures. Indeed, we shall see in section~\ref{sec:mgb2}
that these two pictures, and a plausible but uncontrolled linear stability
analysis, give {\em different} predictions for the anisotropy in the most
important immediate application, magnesium diboride.

\begin{figure}[!ht]
\centering
\includegraphics[width=0.9\linewidth]{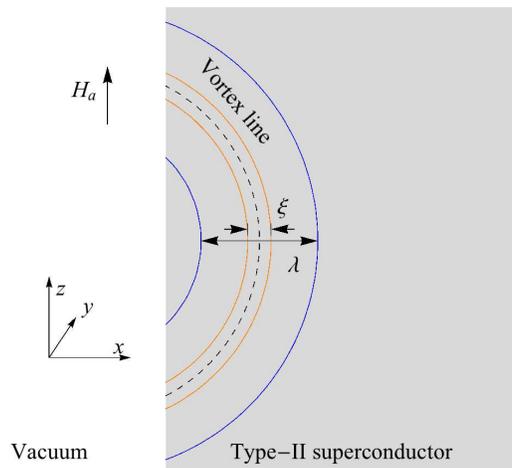}
\caption{(Color online) Illustrating vortex nucleation in a type-II superconductor occupying the half space $x>0$, and subject to a magnetic field parallel to the vacuum-superconductor interface.%
\label{fig:vortexNucleation}}
\end{figure}

Let the superconductor occupy the half space $x>0$, and the magnetic field $\bm{H}$ be parallel to the $z$ axis. Figure~\ref{fig:vortexNucleation} illustrates vortex nucleation in a type-II superconductor for this configuration. With this choice for the system geometry, we neglect effects of field bending over sample corners, which can play a very important role in the flux penetration of real samples. However, we note that these effects are not appreciable for RF cavities for particle accelerators, which have an approximate cylindrical shape in the regions of high magnetic fields.

Let us start with a heuristic estimate of the superheating field for type I superconductors. At an interface between superconductor and insulator (or vacuum), the order parameter $\psi$ is not suppressed; however, if we force a slab of magnetic field into the superconductor thick enough to force the surface to go normal and $\psi\rightarrow 0$, the superconductivity will be destroyed over a depth $\xi$, the coherence length of the SC, with energy cost per unit area $[{H_c}^2 / (8\pi)] \xi_i$, with $i=a$ and $i=c$, for $\bm{c}\parallel z $ and $\bm{c} \perp z$, respectively. The necessary width of the magnetic slab should be set by the Meissner magnetic penetration depth $\lambda$, with approximate energy gain per unit area, given by the magnetic pressure times the depth, or: $[H_{\text{sh}} / (4\pi)] (H_{\text{sh}} \lambda_i)$. Thus $H_{\text{sh}} / (\sqrt{2}H_c)  \approx (1/2) (\lambda_i / \xi_i)^{-1/2} = (1/2) {\kappa_i}^{-1/2}$, which is close to the exact result: $H_{\text{sh}} / (\sqrt{2}H_c)(\kappa \ll 1)  = 2^{-3/4} \kappa^{-1/2}$ for isotropic Fermi surfaces~\cite{transtrum11}. The anisotropy of the superheating field is then proportional to $\gamma^{1/2}$, assuming $\kappa \ll 1$ for $\bm{c}$ parallel and perpendicular to the magnetic field.

\begin{figure}[!ht]
(a) \par\smallskip
\centering
\includegraphics[width=\linewidth]{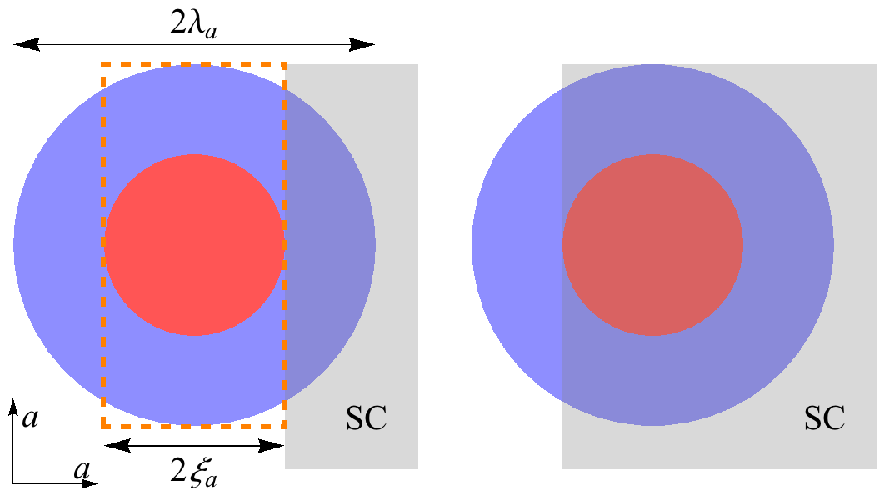}
\\
(b) \par\smallskip
\centering
\includegraphics[width=\linewidth]{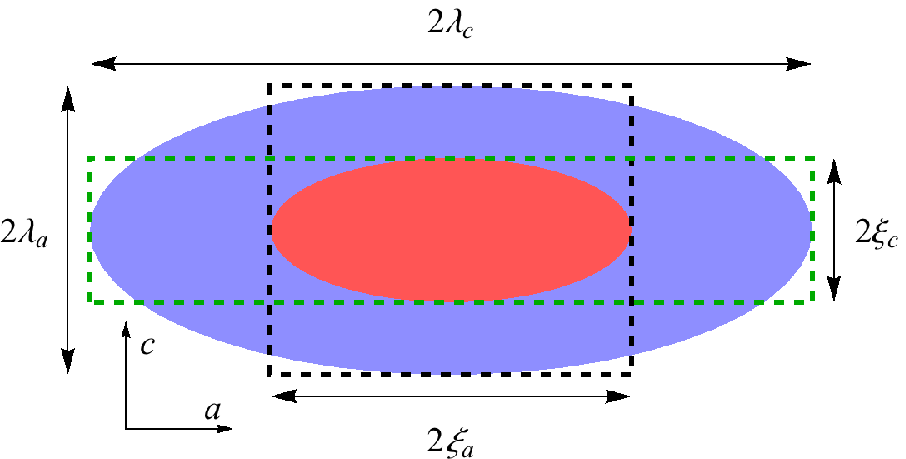}
\caption{(Color online) (a) Illustration of the penetration of a vortex core into a type-II anisotropic superconductor with anisotropy axis $\bm{c} \parallel z$ (perpendicular to the plane of the figure). (b) Vortex and vortex core acquire an ellipsoidal shape when $\bm{c}$ lies in the $xy$ plane. Here the superconductor surface lies horizontally and vertically, for $\bm{c}$ parallel to $x$ and $y$, respectively. The magnetic field is parallel to $z$ for both (a) and (b). We can estimate the superheating field from the calculation of the work necessary to push a vortex core into the superconductor, thus destroying the Meissner state. For anisotropic vortices, the superheating field turns out to be proportional to the area of the green (black) boxes for the superconductor boundary surface parallel (perpendicular) to the $\bm{c}$ axis. These estimates simplify the calculations of Bean and Levingston~\cite{bean64}, which consider the vanishing of the surface energy barrier felt by a single penetrating vortex in a type-II superconductor. More generally, the Ginzburg-Landau approach takes into account the cooperative effects due to the penetration of multiple vortices~\cite{transtrum11}. %
\label{fig:HshEstimation}}
\end{figure}

For type II superconductors, consider the penetration of a vortex core into the superconductor, as illustrated in FIG.~\ref{fig:HshEstimation}a. The vortex and vortex core correspond to the blue and red regions, respectively. The magnetic field $\bm{H}$ is again parallel to $z$ (perpendicular to the plane of the figure), and the anisotropy axis $\bm{c}$ is either parallel (FIG.~\ref{fig:HshEstimation}a) or perpendicular (FIG.~\ref{fig:HshEstimation}b) to $z$; the gray region in FIG.~\ref{fig:HshEstimation}a illustrates a superconductor occupying the semi-infinite space $x>0$. Vortex and vortex core acquire an ellipsoidal shape when $\bm{c}$ lies in the $xy$ plane (FIG.~\ref{fig:HshEstimation}b); here the superconductor surface lies horizontally and vertically when $\bm{c} \parallel x$ and $\bm{c}\parallel y$, respectively. We can estimate the superheating field by comparing the work (per unit length) that is necessary to push a vortex core into the superconductor (thus destroying the Meissner state) with the condensation energy:
\begin{eqnarray}
\frac{H_{\text{sh}}}{4\pi} \Delta H \approx \frac{{H_c}^2}{8\pi} S_{\text{vc}},
\label{eq:hshtype2_eq1a}
\end{eqnarray}
where $S_{\text{vc}}$ is the area of the vortex core (red region in FIG.~\ref{fig:HshEstimation}), and $\Delta H$ is given by
\begin{eqnarray}
\Delta H =  \frac{\Phi_0}{S_{\text{v}}} S_{\Delta},
\label{eq:hshtype2_eq1b}
\end{eqnarray}
where $\Phi_0$ is the fluxoid quantum~\cite{tinkham96}. $S_{\text{v}}$ is the total sectional area of the vortex; e.g. $S_{\text{v}} = \pi \, \lambda^2$ for isotropic superconductors. $S_\Delta$ is the amount of vortex area that penetrates when the vortex core is pushed into the superconductor; it is approximately equal to the areas of the green, black, and orange dashed rectangles in FIG.~\ref{fig:HshEstimation}, for $\bm{c} \parallel x$, $y$ and $z$, respectively. Table~\ref{tab:vortices} shows equations for $S_{\text{v}}$, $S_{\text{vc}}$ and $S_\Delta$ in terms of the penetration and coherence lengths, with $\bm{c}$ parallel to each cartesian axis. Equations (\ref{eq:hshtype2_eq1a}-\ref{eq:hshtype2_eq1b}) then read:
\begin{eqnarray}
H_{\text{sh}} = \frac{{H_c}^2 \pi^2}{8 \, \Phi_0} \times
\begin{cases}
\lambda_c \, \xi_c, & \text{if } \bm{c} \parallel y, \\
\lambda_a \, \xi_a, & \text{if } \bm{c} \parallel x \text{ or } z.
\end{cases}
\label{eq:estimateHsh}
\end{eqnarray}
Interestingly, for $\bm{c} \parallel y$, the penetrating vortex area is the area of the \emph{black} dashed box (see FIG.~\ref{fig:HshEstimation}b), whereas the superheating field is proportional to the area of the dashed \emph{green} box. Conversely, for $\bm{c} \parallel x$, the penetrating vortex area is the area of the \emph{green} dashed box, whereas the superheating field is proportional to the area of the dashed \emph{black} box. Within GL theory, $\lambda_a \, \xi_a = \lambda_c \, \xi_c$, suggesting that the superheating field is isotropic. Plugging $\Phi_0=2\,\sqrt{2}\, \pi H_c \, \lambda_i \, \xi_i$ into Eq. \eqref{eq:estimateHsh}, we find $H_{\text{sh}} / (\sqrt{2} H_c) \approx 0.1$, which is independent of $\kappa$, as in the exact calculations for isotropic Fermi surfaces~\cite{transtrum11}, but off by an overall factor of five from the linear stability results: $H_{\text{sh}} / (\sqrt{2} H_c) (\kappa \gg 1)\approx 0.5$. In section~\ref{sec:gl}, we show that $H_{\text{sh}}$ is isotropic within GL for $\kappa \gg 1$. In section~\ref{sec:mgb2}, we discuss recent work at lower temperatures using the two-band model for MgB${}_2$, which then suggests a substantial anisotropy.

\begin{table}[h]
\centering
\begin{tabular}{ | l | c | c | c | }
\hline
 & $\bm{c} \parallel x$ &  $\bm{c} \parallel y$ & $\bm{c} \parallel z$ \\ \hline
$S_{\text{v}}$ & $\pi {\lambda_a} \lambda_c$ & $\pi {\lambda_a} \lambda_c$ & $\pi {\lambda_a}^2$ \\
$S_{\text{vc}}$ & $\pi {\xi_a} \xi_c$ & $\pi {\xi_a} \xi_c$ & $\pi {\xi_a}^2$ \\
$S_{\Delta}$ & $4 \, \lambda_c \, \xi_c$ & $4 \, \lambda_a \, \xi_a$ & $4 \, \lambda_a \, \xi_a$ \\
\hline
\end{tabular}
\footnotesize
\caption{Area of the vortex ($S_{\text{v}}$), area of the vortex core ($S_{\text{vc}}$) and approximated penetrating field area ($S_{\Delta}$; area of the dashed rectangles in FIG.~\ref{fig:HshEstimation}) for $\bm{c}$ parallel to each cartesian axis.}
\label{tab:vortices}
\end{table}

After the vortex core penetrates the superconductor, the vortex is
subject to an attractive force toward the interface due to the
boundary condition (there is no normal current at the surface).
Bean and Livingston~\cite{bean64} used this to give a second
simple, intuitive estimate of the superheating field.
They model this force as an interaction with an `image vortex' of
opposite sign outside the superconductor, starting the vortex center (somewhat
arbitrarily) at a distance $x=\xi$ from the interface -- precisely where
our estimate left the vortex.  The superheating field is set by the competition
between magnetic pressure and the attractive long-range force.
This leads to the equation
\begin{equation}
H_\text{sh} = \frac{\Phi_0}{4 \pi} \frac{1}{ \lambda \, \xi}.
\label{eq:BLestimate}
\end{equation}
Using the GL relation: $\Phi_0 = 2 \sqrt{2} \, \pi \, H_c \, \lambda \, \xi$,
one finds $H_{\text{sh}} \approx 0.71 H_c$. How can we incorporate
crystal anisotropy into this simple calculation? If vortex and vortex core
have the same shape, we can use Eq.~\eqref{eq:changeVariables} to map
the anisotropic system into an isotropic one with $\xi_y$ and
$\lambda_x$ replacing $\xi$ and $\lambda$. This mapping preserves
magnetic fields, but not loop areas in the $xy$ plane, so that the
fluxoid quantum $\Phi_0$ rescales to $\tilde{\Phi}_0 = (\xi_y/\xi_x)
\Phi_0$ under this change of coordinates. Thus,
$H_{\text{sh}}=\tilde{\Phi}_0/(4 \pi \lambda_x \, \xi_y) = \Phi_0/(4 \pi
\lambda_x \, \xi_x) \approx 0.71 H_c$, which is isotropic and compatible
with the first simple argument, and the results in the next section for
the large $\kappa$ limit of the anisotropic GL theory.

It is interesting and convenient that these two fields (condensation
energy associated with vortex core nucleation and attractive force due
to the boundary conditions) are of the same scale. Bean and Livingston's
estimate results in $H_{\text{sh}}/ H_c = 0.71$, of the same form as our
estimate but larger and closer to the true GL calculation
$H_{\text{sh}}/ H_c = 0.75$. However, we should mention that while the
{\em field} needed to push the core into the superconductor is close to
that needed to push the vortex past the attractive force towards the
`image-vortex', the two contributions contribute very differently to the
energy barrier. Bean and Livingston's force can act on a scale longer by a
factor $\kappa = \lambda/\xi$ than our core nucleation, and will
dominate the barrier height for $H$ near $H_{c1}$.

How is GL different from these two simple pictures? First, the GL calculation
incorporates both the initial core penetration and the long-range attractive 
force. Second, it accounts for the cooperative effects of multiple vortices 
entering at the same time. Third, and perhaps most important, the physical
picture near $H_{\text{sh}}$ is quite different. As discussed
in~\cite{transtrum11}, the wavelength of the sinusoidal instability within
GL theory is $2\pi k_c \propto \kappa^{1/4} \xi$. The single vortex within
our model and Bean and Livingston have sharp cores of size $\xi$; the correct
linear-stability result has the superconducting order parameter varying
smoothly over a longer length larger by $\kappa^{1/4}$. We shall see in 
section~\ref{sec:mgb2} that taking these three basic methods outside
the realm where GL theory is valid yields three quite different predictions
for the anisotropy in the superheating field.

\section{Ginzburg-Landau theory of the superheating field anisotropy}
\label{sec:gl}

Let us flesh out these intuitive limits into a full calculation. A phenomenological generalization of GL theory that incorporates the anisotropy of the Fermi surface was initially proposed by Ginzburg~\cite{ginzburg52}, and revisited later, using the microscopic theory, by several authors~\cite{caroli63, gorkov64, tilley65}. In this approach, the gauge-invariant derivative terms are multiplied by an anisotropic effective mass \emph{tensor} that depends on integrals over the Fermi surface (see e.g. Eq. 2 of Ref.~\cite{tilley65}). The mass tensor is a multiple of the identity matrix for cubic crystals, such as Nb, Nb${}_3$Sn and NbN, which belong to the next generation of superconducting accelerator cavities. In this case, the dominant effects of the Fermi surface anisotropy are higher-order multipoles, which may be added using, e.g. nonlocal terms of higher gradients~\cite{hohenberg67}. On the other hand, as it should be anticipated, mass anisotropy can lead to important effects on layered superconductors, such as MgB${}_2$ and some iron-based superconductors (also considered for RF cavities), at least insofar as the GL formalism is accurate. Simple arguments within GL theory can be used to show that the anisotropy of the upper-critical and lower-critical fields is proportional to $\gamma$; i.e. $H_{c2}^\perp / H_{c2}^\parallel = \gamma = H_{c1}^\parallel / H_{c1}^\perp$, where the perpendicular (parallel) symbol indicates that the applied magnetic field is perpendicular (parallel) to the $\bm{c}$ axis. The effects of Fermi surface anisotropy on the properties of superconductors have been theoretically studied by many authors~\cite{ginzburg52, gorkov64, tilley65, hohenberg67, daams81, kogan03}.

One possible generalization of the Ginzburg-Landau free energy to incorporate anisotropy effects has been written down by Tilley~\cite{tilley65}:
\begin{eqnarray}
f_s - f_n &=& \sum_{i,j \in \{x,y,z\}} \frac{1}{2\,m_{ij}} \left(-\frac{\hbar}{i} \frac{\partial \psi^*}{\partial x_i} - \frac{e^*}{c} A_i \psi^* \right) 
\nonumber \\ && \quad
\times \left(\frac{\hbar}{i} \frac{\partial \psi}{\partial x_j} - \frac{e^*}{c} A_j \psi \right)+ \alpha |\psi|^2 + \frac{\beta}{2} |\psi|^4
\nonumber \\ && \quad
+ \frac{\left(\bm{H}_a-\nabla \times \bm{A} \right)^2}{8 \pi},
\label{free_energy_1}
\end{eqnarray}
where $f_s$ and $f_n$ are the free energy densities of the superconducting and normal phases, respectively; $\psi$ is the superconductor order parameter, $\bm{A}$ is the vector potential, and $\bm{H}_a$ is an applied magnetic field. Anisotropy is incorporated in the effective mass tensor $M = ((m_{ij}))$, whose components can be conveniently expressed as a ratio of integrals over the Fermi surface (see Eq. (2) of Ref.~\cite{tilley65}). $e^*$ is the effective charge, $\alpha$ and $\beta$ are energy constants, and $\hbar$ and $c$ are Plank's constant (divided by $2\pi$) and the speed of light, respectively. The thermodynamic critical field is given by~\cite{tinkham96}: $H_c = \sqrt{4 \pi \alpha^2 / \beta}$, independent of mass anisotropy. Eq. \eqref{free_energy_1} can then be written in a more convenient form:
\begin{eqnarray}
\frac{\left(f_s - f_n \right)}{{H_c}^2/(4\pi)} &=& \sum_i \left[\left(\xi_i\frac{\partial f}{\partial x_i}\right)^2 + \left(\xi_i \frac{\partial \phi}{\partial x_i} - \frac{A_i}{\sqrt{2} H_c \lambda_i} \right)^2 f^2 \right]
\nonumber \\ && \quad
+ \frac{1}{2} \left(1- f^2\right)^2 + \frac{1}{2{H_c}^2}\left(\bm{H}_a - \nabla \times \bm{A} \right)^2,
\label{free_energy_2}
\end{eqnarray}
where we have assumed a layered superconductor with the anisotropy axis $\bm{c}$ aligned with one of the three Cartesian axes, so that $i\in \{x,y,z\}$ in the first term of the right-hand side, and we have dropped an irrelevant additive constant $1/2$. Also, we have rewritten the order parameter as $\psi = |\psi_\infty| \, f \, e^{i \phi}$, where $f$ and $\phi$ are scalar fields, and $\psi_\infty=-\alpha/\beta$ is the solution infinitely deep in the interior of the superconductor~\cite{tinkham96}. The anisotropic penetration depth and coherence lengths are given by $\lambda_i = (m_i c^2 / (4\pi |\psi_\infty|^2 {e^*}^2) )^{1/2}$, and $\xi_i = ( \hbar^2 / ( 2m_i (-\alpha) ) )^{1/2}$, respectively.

\begin{figure}[!ht]
\centering
\includegraphics[width=\linewidth]{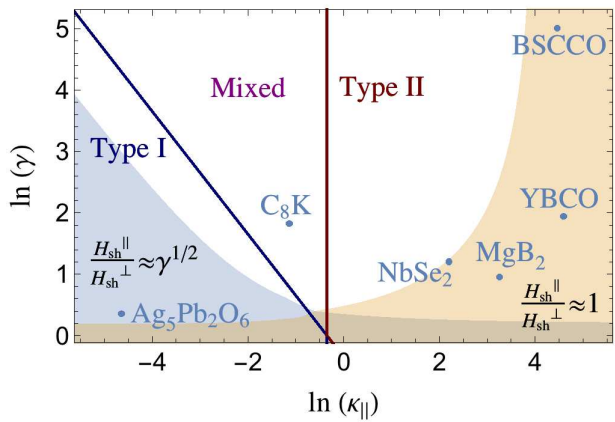}
\caption{(Color online) Showing regions in $\kappa_\parallel \times \gamma$ space where the crystal is always type I (left region to blue solid lines), always type II (right region to red solid lines), or might be of either type (region between red and blue lines), depending on the orientation of the crystal. The shaded blue and orange regions correspond to regions where the Ginzburg-Landau superheating field anisotropy can be approximated by $\gamma^{1/2}$ and $1$, respectively, within $10\%$ of accuracy. %
\label{fig:kappaxgammaDiagram}}
\end{figure}

Let the pairs of characteristic lengths $(\lambda_c, \xi_c )$ and $(\lambda_a, \xi_a )$ be associated with the layer-normal and an in-plane axis, respectively. Define:
\begin{eqnarray}
\kappa_\parallel \equiv \frac{\lambda_a}{\xi_a}, \quad \kappa_\perp \equiv \frac{\lambda_c}{\xi_a} = \frac{\lambda_a}{\xi_c} = \gamma \, \kappa_\parallel,
\end{eqnarray}
where the last two relations can be verified using the definition of $\lambda_i$, $\xi_i$, and $\gamma$. Following previous calculations of the superheating field~\cite{catelani08, transtrum11}, we also let $\bm{H}_a$ be parallel to $z$, and the superconductor occupy the half-space region $x>0$, so that symmetry constraints imply that $A_z=0$, and all fields should be independent of $z$. Thus, if the anisotropy axis $\bm{c}$ is parallel to $z$, our GL free energy (Eq. \ref{free_energy_2}) is directly mapped into the isotropic free energy of Transtrum et al.~\cite{transtrum11}, with $\xi$ and $\lambda$ replaced by $\xi_a$ and $\lambda_a$, respectively. In particular, the solution for the superheating field $H_{\text{sh}}$ as a function of $\kappa$ is given in Ref.~\cite{transtrum11} using $\kappa_\parallel$ instead of $\kappa$. If $\bm{c}$ is parallel to $x$ or $y$, there are a number of scaling arguments that can be used to map the anisotropic free energy into the isotropic one~\footnote{For instance, Blatter et al. recognized that by making a change of coordinates and redefining the magnetic field and vector potential, one could make isotropic the derivative term by introducing anisotropy in the magnetic energy terms~\cite{tinkham96, blatter92}.}. Here we consider the change of coordinates and rescaling of the vector potential:
\begin{eqnarray}
\bm{r} = \left(\frac{\xi_x}{\xi_y} \tilde{x}, \tilde{y}, \tilde{z} \right),
\quad
\bm{A}=\left( \tilde{A}_x, \frac{\xi_x}{\xi_y} \tilde{A}_y, \tilde{A}_z\right).
\label{eq:changeVariables}
\end{eqnarray}
Note that this change of variables does not change the magnetic field, since $\bm{H}_a$ is aligned with the $z$ axis, so that the $z$-component of the field is given by: $\partial A_y / \partial x - \partial A_x / \partial y = \partial \tilde{A}_y / \partial \tilde{x} - \partial \tilde{A}_x / \partial \tilde{y}$. This coordinate transformation maps the anisotropic free energy into an isotropic one with $\xi_y$ and $\lambda_x$ replacing $\xi$ and $\lambda$. In particular, now the solution for the superheating field is given in Ref.~\cite{transtrum11} using $\kappa_\perp=\gamma \kappa_\parallel$ instead of $\kappa$. In this paper, we only consider the two representative cases: $\bm{c} \parallel z$ and $\bm{c} \perp z$, as we do not expect appreciable qualitative changes for arbitrary orientations of $\bm{c}$ with respect to the $z$ axis. Notice the interesting fact that a crystal might be a type I superconductor ($\kappa_\parallel < 1 / \sqrt{2}$) when $\bm{c}$ is parallel to $z$, and yet be a type II superconductor if $\gamma \kappa_\parallel> 1/ \sqrt{2}$ when $\bm{c}$ is perpendicular to $z$ (see Fig.~\ref{fig:kappaxgammaDiagram}). This interesting property of anisotropic superconductors has been discussed in Ref.~\cite{kogan14}, and confirmed experimentally in the work of Ref.~\cite{koike80}.

\begin{figure}[!ht]
\centering
\includegraphics[width=0.9\linewidth]{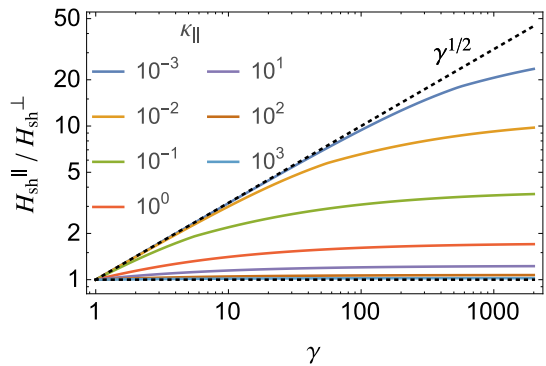}
\caption{(Color online) Anisotropy of the Ginzburg-Landau superheating field as a function of mass anisotropy for several values of $\kappa_\parallel$. The dotted lines are the limiting solutions given by Eq.~\ref{eq:limitAnisotropy}.%
\label{fig:gammash}}
\end{figure}

Now we turn our attention to the anisotropy of the superheating field:
\begin{eqnarray}
\frac{H_{\text{sh}}^\parallel}{H_{\text{sh}}^\perp}=\frac{H_{\text{sh}} (\kappa_\parallel)}{H_{\text{sh}} (\kappa_\perp)}=\frac{H_{\text{sh}} (\kappa_\parallel)}{H_{\text{sh}} (\gamma \, \kappa_\parallel)}.
\label{eq:anisotropyHsh}
\end{eqnarray}
For general $\kappa$, approximate solutions for the superheating field for isotropic systems are given by Eqs. (10) and (11) of Ref.~\cite{transtrum11}, which we reproduce here for convenience:
\begin{eqnarray}
\frac{H_{\text{sh}}}{\sqrt{2}H_c} \approx
2^{-3/4} \kappa^{-1/2} \frac{1+ 4.6825120 \, \kappa + 3.3478315 \, \kappa^2 }{1+ 4.0195994 \, \kappa + 1.0005712 \, \kappa^2}, 
\label{eq:asymp_small}
\end{eqnarray}
for small $\kappa$, and
\begin{eqnarray}
\frac{H_{\text{sh}}}{\sqrt{2}H_c} \approx
\frac{\sqrt{10}}{6} + 0.3852 \, \kappa^{-1/2},
\label{eq:asymp_large}
\end{eqnarray}
for large $\kappa$. We can use approximations \eqref{eq:asymp_small} and \eqref{eq:asymp_large} to find asymptotic solutions for the superheating field anisotropy:
\begin{eqnarray}
\frac{H_{\text{sh}}^\parallel}{H_{\text{sh}}^\perp} \approx
\left\{\begin{array}{ll}
\gamma^{1/2}, & \text{for } \kappa \ll 1/\gamma, \\
1, & \text{for } \kappa \gg 1,
\end{array}
\right.
\label{eq:limitAnisotropy}
\end{eqnarray}
with $\gamma>1$. These asymptotic solutions span a large region in the phase diagram of Fig. \ref{fig:kappaxgammaDiagram}, with the shaded blue and orange regions corresponding to regions where the superheating field anisotropy can be approximated by $\gamma^{1/2}$ and $1$, respectively. Figure~\ref{fig:gammash} shows a plot of the anisotropy of the superheating field as a function of the mass anisotropy for several values of $\kappa_\parallel$. The dotted lines are asymptotic solutions given by Eq. \eqref{eq:limitAnisotropy}. In order to make this plot we considered the solution for the superheating field to be given by Eqs. \eqref{eq:asymp_small} and \eqref{eq:asymp_large} for $\kappa<\kappa_{\text{th}}$ and $\kappa \ge \kappa_{\text{th}}$, respectively, where the threshold $\kappa_{\text{th}} \approx 0.56$ is found by equating the right-hand sides of the two approximate solutions. It is clear that the combination of large $\gamma$ and small $\kappa$ yields the largest anisotropy of $H_{\text{sh}}$. Notice that the deviation from the simple asymptotic solution at small $\kappa_\parallel$ scales as $(H_{\text{sh}}^\parallel / H_{\text{sh}}^\perp - \gamma^{1/2}) / \gamma^{1/2} = \mathcal{O} (\kappa \gamma)$.

\begin{table}[h]
\centering
\begin{tabular}{ | l | c | c | c | }
\hline
\text{Material} & $\kappa_\parallel$ &  $\gamma$ &  $H_{\text{sh}}^{\parallel}/H_{\text{sh}}^{\perp}$ \\ \hline
Ag${}_5$Pb${}_2$O${}_6$ (Ref.~\cite{mann07}) & $\sim 0.0096$ & $\sim 1.43$ & $\sim 1.2$ \\
C${}_8$K (Ref.~\cite{koike80}) & $\sim 0.32$ & $\sim 6.2$ & $\sim 1.6$ \\
NbSe${}_2$ (Ref.~\cite{trey73}) & $\sim 9$ & $\sim 3.33$ & $\sim 1.1$ \\
MgB${}_2$ (Refs.~\cite{chen01, kogan03}) & $\sim 26$ & $\sim 2.6$ & $\sim 1.05$ \\
BSCCO (Refs.~\cite{stintzing97, tinkham96}) & $\sim 87$ & $\sim 150$ & $\sim 1.07$ \\
YBCO (Refs.~\cite{stintzing97, tinkham96}) & $\sim 99$ & $\sim 7$ & $\sim 1.04$ \\
\hline
\end{tabular}
\footnotesize
\caption{Ginzburg-Landau parameter $\kappa_\parallel$ with $\bm{c}$ parallel to the $z$ axis, mass anisotropy $\gamma$, and superheating field anisotropy for different materials.}
\label{tab:materialAnistropies}
\end{table}

\begin{figure}[!h]
\centering
\includegraphics[width=0.9\linewidth]{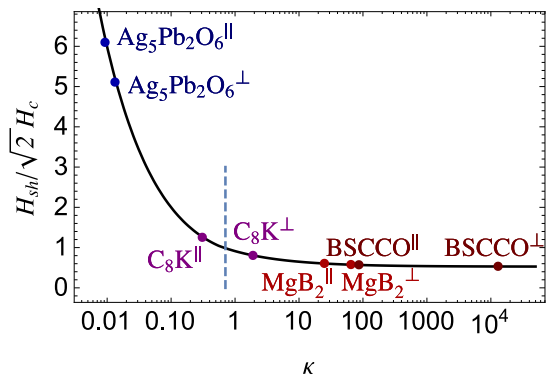}
\caption{(Color online) Ginzburg-Landau superheating field $H_{\text{sh}}/(\sqrt{2}H_c)$ as a function of $\kappa$. The points correspond to the solutions using $\kappa_\parallel$ and $\kappa_\perp$ for Ag${}_5$Pb${}_2$O${}_6$ (blue), C${}_8$K (purple), MgB${}_2$ (red), BSCCO (dark red).\label{fig:hsh}}
\end{figure}

In Table~\ref{tab:materialAnistropies} we compare the anisotropy of the superheating field for different materials; we also present the values that we used for $\kappa_{\parallel}$ and $\gamma$ in each case. As we have stressed before, the superheating field anisotropy is largest for small $\kappa_\parallel$ and large $\gamma$. Note that even though type-I superconductors have small $\kappa$, we have not found anisotropy parameters for elemental superconductors in the literature, probably because anisotropy plays a minor role for most of them. Just a few well-studied non-elemental superconductors are of type I, such as the layered silver oxide Ag${}_5$Pb${}_2$O${}_6$, with a mass anisotropy of about $1.43$, and $\kappa_\parallel \approx 0.01 < 1/\sqrt{2} $. On the other hand, type-II superconductors are known for their large anisotropies. The critical fields of BSCCO, for instance, can vary by two orders of magnitude depending on the orientation of the crystal. Yet the anisotropy effects on the superheating field are undermined (Eq. \eqref{eq:anisotropyHsh}) by the flat behavior of $H_{\text{sh}}$ at large $\kappa$. These effects are also illustrated in Fig.~\ref{fig:hsh}, where we plot the solution $H_{\text{sh}}/(\sqrt{2}H_c)$ as a function of $\kappa$, using the asymptotic solutions given by Eqs. \eqref{eq:asymp_small} and \eqref{eq:asymp_large} for $\kappa<=0.56$ and $\kappa>0.56$, respectively (this approximated solution is remarkably close to the exact result~\cite{transtrum11}). Note that within GL theory $H_{\text{sh}}/H_c$ depends on material properties only through the parameter $\kappa$. The points in FIG. \ref{fig:hsh} correspond to the solutions of the superheating field using $\kappa_\parallel$ and $\kappa_\perp$ for Ag${}_5$Pb${}_2$O${}_6$ (blue), C$_{8}$K (purple), MgB${}_2$ (red), BSCCO (dark red). Superconductors with $\kappa_\parallel \approx 1$ can have an enormous anisotropy $\gamma$, say $\sim 10^{5}$, and yet the superheating field will be nearly isotropic.

One should bear in mind that GL formalism is accurate only in the narrow ranges of temperatures near the critical point. Beyond this range, one must rely either on generalizations of GL to arbitrary temperatures~\cite{tewordt63, werthamer63}, or more complex approaches using BCS theory, Eilenberger semi-classical approximation and strong-coupling Eliashberg theory. However, note that GL and Eilenberger theories yield similar quantitative results for the temperature dependence of the superheating in the limit of large $\kappa$ for isotropic Fermi surfaces (see e.g. Ref.~\cite{catelani08}).

\section{Low-temperature anisotropy of the superheating field for MgB${}_2$}
\label{sec:mgb2}

We now turn to MgB$_2$, an anisotropic, layered superconductor which would 
likely in practice be used at temperatures $T\ll T_c$ where GL theory is not
a controlled approximation. Here we discover that we get three rather
different estimates for the anisotropy in the superheating field, from our
two simple estimates of section~\ref{sec:estimations} and from an uncontrolled GL-like
linear stability analysis.

\begin{figure}[!h]
\centering
\includegraphics[width=0.6\linewidth]{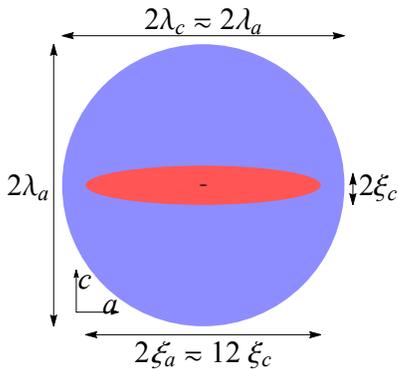}
\caption{(Color online) Illustrating a vortex and vortex core (blue and red regions) of MgB${}_2$ near $T=0$ (we increased $\xi_a$ by a factor of $30$ with respect to $\lambda_a$, so that the core features become discernible; the small black region in the center corresponds to the actual scale). Near zero temperature, the field penetration region is calculated to be nearly isotropic ($\lambda_a \approx \lambda_c$), whereas the core shape anisotropy is predicted to reach a maximum ($\xi_a \approx 6\, \xi_c$) (Ref.~\cite{kogan03}).
\label{fig:lowTmgb2}}
\end{figure}

The striking qualitative difference between low temperature MgB$_2$ and
GL theory is the violation of the GL anisotropy relation:
$\lambda_c / \lambda_a \neq \xi_a / \xi_c$. For anisotropic superconductors, this originates in the mass dependence of the penetration and coherence lengths ($\lambda \sim m^{1/2}$ whereas $\xi \sim m^{-1/2}$). Experiments~\cite{angst02,budko02,cubitt03a,cubitt03b,lyard04,budko15} and theoretical calculations~\cite{kogan02a,kogan02b,kogan03} for MgB${}_2$ suggest that this relation is violated at lower temperatures; the anisotropies of $\lambda$ and $\xi$ exhibit opposite temperature dependences, with
\begin{eqnarray}
\gamma_\lambda = \lambda_c / \lambda_a
\end{eqnarray}
increasing, whereas
\begin{eqnarray}
\gamma_\xi = \xi_a / \xi_c
\end{eqnarray}
decreases~\footnote{Here we assume that the anisotropy $\gamma_\xi = \xi_a / \xi_c$ is equivalent to the anisotropy of the upper-critical field $\gamma_H = H_{c2,a} / H_{c2,c}$, as in the work of Ref.~\cite{kogan02b}.} with temperature. Figure~\ref{fig:lowTmgb2} shows an illustration of a vortex section near $T=0$. Using calculations from Ref.~\cite{kogan03}, $\gamma_\lambda$ and $\gamma_\xi$ become equal only at $T=T_c$.

We can use our first method to estimate the low-temperature superheating field anisotropy by relaxing the constraint $\lambda_c \, \xi_c = \lambda_a \, \xi_a $ in Eq.~\eqref{eq:estimateHsh} of section \ref{sec:estimations}, resulting,
\begin{eqnarray}
\frac{H_{\text{sh}}^{c \perp y}}{H_{\text{sh}}^{c \parallel y}} = \frac{\gamma_\xi}{\gamma_\lambda},
\label{eq:anisotropyMgB2}
\end{eqnarray}
where $H_{\text{sh}}^{c \perp y}$ means either $H_{\text{sh}}^{c \parallel x}$ or $H_{\text{sh}}^{c \parallel z}$. $H_{\text{sh}}$ is isotropic near $T=T_c$, since $\gamma_\xi \approx \gamma_\lambda$. 

Our other two estimates rely on an uncontrolled approximation---using Eq.~\eqref{free_energy_2} with the low temperature values of $\lambda$ and $\xi$. This is not justified microscopically, as the calculations of section~\ref{sec:gl}.
Our second estimate draws on Bean and Livingston to estimate the anisotropy. 
For the case
$\bm{c}$ in the $xy$ plane and $\lambda_c/\lambda_a \neq \xi_a / \xi_c$, rather than
using Eq.~\eqref{eq:changeVariables}, let us consider the rescaling:
$\bm{r} = ((\lambda_y/\lambda_x)\, \tilde{x}, \tilde{y}, \tilde{z} )$, and $\bm{A} = (\tilde{A}_x, (\lambda_y / \lambda_x) \, \tilde{A}_y, \tilde{A}_z )$. If we plug these equations into Eq.~\eqref{free_energy_2}, assuming $\gamma_\lambda \neq \gamma_\xi$, we would obtain a GL theory that is isotropic in $\lambda$, but anisotropic in $\xi$, with $\lambda \rightarrow \lambda_x$, $\xi_x \rightarrow (\lambda_x / \lambda_y) \xi_x$, and $\Phi_0 \rightarrow \tilde{\Phi}_0 = (\lambda_x/\lambda_y) \Phi_0$. Now we can plug the new lengths and $\tilde{\Phi}_0$ into Bean and Livingston's calculation to obtain:
\begin{align}
H_{\text{sh}} & = \frac{\tilde{\Phi}_0}{4 \pi} \frac{1}{ \lambda_x \, (\lambda_x / \lambda_y) \xi_x} 
= \frac{\Phi_0}{4 \pi}
\begin{cases}
(\lambda_c \, \xi_c)^{-1} & \text{for } \bm{c} \parallel x, \\
(\lambda_a \, \xi_a)^{-1} & \text{for } \bm{c} \parallel y.
\end{cases}
\label{eq:BLaniEstimate}
\end{align}
Note that unlike the GL case, $\Phi_0$ cannot be written as $2 \, \sqrt{2} \, \pi \, H_c \, \lambda \, \xi$, and the superheating field is not isotropic. We find that $H_{\text{sh}}^{c \parallel x} / H_{\text{sh}}^{c \parallel y} = \gamma_\xi / \gamma_\lambda$, as in Eq.~\eqref{eq:anisotropyMgB2}. Unlike our first estimate, where $z$ and $x$ are equivalent directions, in this adaptation of Bean and Livingston's we find that $y$ and $z$ are equivalent directions. On the one hand, the only relevant component of the coherence length is the one that is parallel to the $x$ axis in Bean and Livingston's argument. On the other hand, our estimates assign different energy barriers to vortex core sections with different areas ($\pi \xi_a^2$ for $\bm{c} \parallel z$, and $\pi \xi_a \xi_c$ for $\bm{c} \parallel y$).

Finally, we note that, while the GL free energy of Eq.~\ref{free_energy_1} enforces
the high-temperature anisotropy relation violated by low-temperature MgB$_2$,
when we rewrite it as Eq.~\ref{free_energy_2} we get a legitimate,
albeit uncontrolled, description
of a superconductor with independent anisotropies for $\lambda$ and $\xi$.
A direct numerical calculation using linear stability analysis on Eq.~\ref{free_energy_2} for the parameters of MgB${}_2$ yields an almost isotropic result: $H_{\text{sh}}^{c \parallel x} / H_{\text{sh}}^{c \parallel y} \approx 1$, and $H_{\text{sh}}^{c \parallel z} / H_{\text{sh}}^{c \parallel x} = 1.03$. Analytical calculations in the large-$\kappa$ limit (using the methods developed in the Appendix of Ref.~\cite{transtrum11}) corroborate this result; the anisotropy vanishes in the high $\kappa$ limit of Eq.~\eqref{free_energy_2} for independent $\lambda$ and $\xi$ as well.

What do these estimates suggest for MgB$_2$? Near $T=0$, the theoretical calculations of Ref.~\cite{kogan03} using a two-gap model for MgB${}_2$ suggest that $\gamma_\xi \approx 6$ and $\gamma_\lambda \approx 1$. Experimental results agree with the theoretical predictions near zero temperature, with $\gamma_\lambda$ being almost isotropic~\cite{cubitt03a,cubitt03b}, and $\gamma_\xi \approx 6-7$ (see e.g. Ref.~\cite{budko15}). However, beware that reported experimental results for $\gamma_\xi$ range from $\approx 1$ to $\approx 13$ (see Ref.~\cite{kogan03} and references therein).

\begin{table}[h]
\centering
\begin{tabular}{ | l | c | c | c | c | }
\hline
\multirow{2}{*}{Approach}  & \multicolumn{3}{c |}{ $H_{\text{sh}}$ ( Tesla ) } & \multirow{2}{*}{Max. Anis.} \\ \hhline{~---~}
 & $\bm{c} \parallel \bm{x}$ & $\bm{c} \parallel \bm{y}$ & $\bm{c} \parallel \bm{z}$ & \\ \hline
1st estimate & $0.04$ & $0.006$ & $0.04$ & $\sim 6$ \\
1st (corrected) & $0.2$ & $0.03$ & $0.2$ & $\sim 6$ \\
2nd estimate (B \& L) & $1.13$ & $0.18$ & $0.18$ & $\sim 6$ \\
``GL'' (Eq.~\eqref{free_energy_2}) & $0.21$ & $0.22$ & $0.22$ & $\sim 1$ \\
\hline
\end{tabular}
\footnotesize
\caption{Estimates of the superheating field and maximum anisotropy of low-temperature MgB$_2$ for the three geometries.}
\label{tab:mgb2Tab}
\end{table}

We summarize our estimates of the superheating field for the three geometries in Table~\ref{tab:mgb2Tab}, using $H_c (0) = 0.26 \,\text{T}$ from Ref.~\cite{wang01}. Recall that our first estimates were off from actual GL calculations by a factor of five. We hence multiply $H_{\text{sh}}$ by this factor at lower temperatures, and use this correction to calculate the results displayed on the second row of the table: ``1st (corrected)''. The last row summarizes the results of the last paragraph, and the last column shows the maximum superheating field anisotropy according to the three methods. In comparison, for Nb the superheating field from Ginzburg-Landau theory extrapolated to low temperature is $0.24$ Tesla~\cite{padamsee09}. 

Several things to note about these estimates.
(1)~All three methods suggest that, perhaps with suitable surface alignment,
MgB$_2$ can have superheating fields comparable to current Nb cavities, with 
a much higher transition temperature (and hence much lower Carnot cooling
costs and likely much lower surface resistance). (2)~One of the three
methods suggests that a particular alignment could yield a significantly
higher superheating field than Nb. (3)~It is not a surprise that these
three estimates differ. As discussed in section~\ref{sec:estimations}, 
the three methods have rather different microscopic pictures of the 
superheating instability; the surprise is that they all give roughly the 
same estimate within GL theory. (The further agreement within anisotropic
GL can be understood as a consequence of our coordinate transformation, 
Eq.~\ref{eq:changeVariables}.) 

Before plunging into an intense development
effort for MgB$_2$ cavities, it would be worthwhile to find out whether
there are dangerous surface orientations, or surface orientations that
would provide significant enhancements -- both of which are allowed
by one of our current estimates. Clearly a direct experimental measurement
on oriented single crystal samples would be ideal, although the engineering
challenge of reaching the theoretical maximum superheating field for
a new material could be daunting. Alternatively, it would be challenging
but possible do a more sophisticated theoretical calculation for the
superheating anisotropy.  Eilenberger theory could be solved either
numerically~\cite{transtrum16} or in the high-$\kappa$
limit~\cite{catelani08} to address lower temperatures. Eliashberg
theory~\cite{kortus01,an01,choi02,liu01}, which incorporates realistic modeling of the
two anisotropic gaps and anisotropic electron-phonon couplings, could
be generalized to add a free surface and the resulting system could 
be solved using linear stability analysis.

\section{Concluding Remarks}
\label{sec:conclusions}

To conclude, we used a generalized Ginzburg-Landau approach to investigate the effects of Fermi surface anisotropy on the superheating field of layered superconductors. Using simple scaling arguments, we mapped the anisotropic problem into the isotropic one, which has been previously studied by Transtrum et al.~\cite{transtrum11}, and show that the superheating field anisotropy depends only on two parameters, $\gamma = \lambda_c / \lambda_a$ and $\kappa_\parallel = \lambda_a / \xi_a$. $H_{\text{sh}}^\parallel / H_{\text{sh}}^\perp$ is larger when $\gamma$ is large and $\kappa_\parallel$ is small, and displays the asymptotic behavior $H_{\text{sh}}^\parallel / H_{\text{sh}}^\perp \approx 1$ for $\kappa_\parallel \gg 1 / \gamma$, and $H_{\text{sh}}^\parallel / H_{\text{sh}}^\perp \approx \gamma^{1/2}$, for $\kappa_\parallel \ll 1$, suggesting that the superheating field is typically isotropic for most layered unconventional superconductors, even for very large $\gamma$ (see Table~\ref{tab:materialAnistropies}), when GL is valid. We surmise that the anisotropy of the superheating field is even smaller for cubic crystals, where higher-order and/or non-linear terms have to be included in the GL formalism.

As a practical question, accelerator scientists have explored stamping radio-frequency cavities out of single-crystal samples, to test whether grain boundaries were limiting the performance of particle accelerators. Our study was motivated by the expectation that one could use this expertise to control the surface orientation in the cavity. Such control likely may yield benefits through either optimizing anisotropic surface resistance or optimizing growth morphology, for deposited compound superconductors (growing Nb$_3$Sn from a Sn overlayer). Our calculations suggest that, for the high-$T_c$, high-$\kappa$ materials under consideration for the next generation of superconducting accelerator cavities, that the theoretical bounds for the maximum sustainable fields will not have a significant anisotropy near $T=T_c$. However, the extension of our intuitive arguments for MgB${}_2$ to low temperatures, using results from a two-gap model within BCS theory (Ref.~\cite{kogan03}), suggest a high value for the anisotropy of $H_{\text{sh}}$ near $T=0$, contrasting with the numerical linear stability analysis of Eq.~\eqref{free_energy_2} using the parameters for low-temperature MgB$_2$, which suggest that the superheating field is still isotropic. This motivates further investigations by means of more sophisticated approaches and experiments controlling surface orientation.

\begin{acknowledgments}
We would like to thank G. Catelani, M. Liepe, S. Posen, and J. She for useful conversations. This work was supported by the U.S. National Science Foundation under Award OIA-1549132, the Center for Bright Beams, and the Grant No. DMR-1312160.
\end{acknowledgments}

%

\end{document}